\documentclass[prl,twocolumn,showpacs,preprintnumbers,amsmath,amssymb,superscriptaddress,nofootinbib,english]{revtex4-1}
\usepackage{graphicx,natbib,times}
\usepackage{url}
\usepackage{color}
\usepackage{rotating}
\usepackage{hyperref}
\hypersetup{
    colorlinks=true,
    linkcolor=red,
    citecolor=blue,
}

\newcommand{\red}[1]{\textcolor{black}{#1}}

\def\ie{{\frenchspacing\it i.e.}}

\def\be{\begin{equation}}
\def\ee{\end{equation}}
\def\ba{\begin{eqnarray}}
\def\ea{\end{eqnarray}}
\def\gsim{\;\rlap{\lower 2.5pt
\hbox{$\sim$}}\raise 1.5pt\hbox{$>$}\;}
\def\lsim{\;\rlap{\lower 2.5pt
\hbox{$\sim$}}\raise 1.5pt\hbox{$<$}\;}

\begin{document}

\title{New probe of departures from general relativity using Minkowski functionals}

\author{Wenjuan Fang}
\email{Corresponding author.\\wjfang@ustc.edu.cn}
\affiliation{Department of Astronomy, University of Science and Technology of China, Hefei, Anhui, 230026, P. R. China}
\affiliation{Key Laboratory for Research in Galaxies and Cosmology, Chinese Academy of Sciences, Hefei, Anhui, 230026, P. R. China}

\author{Baojiu Li}
\email{baojiu.li@durham.ac.uk}

\affiliation{Institute for Computational Cosmology, Department of Physics, Durham University, Durham, DH1 3LE, UK}

\author{Gong-Bo Zhao}
\email{gbzhao@nao.cas.cn}

\affiliation{National Astronomy Observatories,
Chinese Academy of Science, Beijing, 100012, P. R. China}
\affiliation{Institute of Cosmology and Gravitation, University of Portsmouth,
Portsmouth, PO1 3FX, UK}

\begin{abstract}

The morphological properties of large scale structure of the Universe can be fully described by four Minkowski functionals (MFs), which provide important complementary information to other statistical observables such as the widely used 2-point statistics in configuration and Fourier spaces. In this work, for the first time, we present the differences in the morphology of large scale structure caused by modifications to general relativity (to address the cosmic acceleration problem), by measuring the MFs from N-body simulations of modified gravity and general relativity. We find strong statistical power when using the MFs to constrain modified theories of gravity: with a galaxy survey that has survey volume $\sim 0.125 (h^{-1}$Gpc$)^3$ and galaxy number density $\sim 1 / (h^{-1}$Mpc$)^{3}$, the two normal-branch DGP models and the F5 $f(R)$ model that we simulated can be discriminated from $\Lambda$CDM at a significance level $\gsim 5\sigma$ with an individual MF measurement. Therefore, the MF of large scale structure is potentially a powerful probe of gravity, and its application {\red{to}} real data deserves active explorations.

\end{abstract}

\pacs{ }

\maketitle

{\it Introduction.} Gravity is one of the most fundamental forces that shape our world, and its effect is felt from very small scales -- in our everyday life -- to very large scales -- in the evolution of our Universe. The year of 2015 celebrated the centenary of our standard theory of gravity -- Einstein's General Relativity (GR) -- which has withstood various rigorous tests. Less attention, however, has been paid to the unsettling fact that these tests were primarily carried out in small systems such as our Solar system, and the application {\red{to}} cosmology is an extrapolation dramatically outside the regime where the theory is confirmed experimentally. The discovery of the accelerating Hubble expansion in the late Universe \cite{perlmutter1998, riess1998}, indeed, lends support to the suspicion that gravity may not be strictly Einsteinian on cosmic scales. Studies of cosmological tests of gravity, therefore, can address these two fundamental questions simultaneously.

There has been a growing body of recent research on testing gravity in cosmology. Most of these studies employ traditional measurements of the geometry and structure formation of the Universe, such as the type Ia supernovae, baryonic acoustic oscillations, gravitational lensing, galaxy clustering and clusters of galaxies. These observables have their advantages and disadvantages, and are usually complementary to each other. The bottom line, however, is that the real Universe has a very complicated structure, which can rarely be fully described by a single observable. Therefore, it is critical that other statistical properties of the same observations, which encode additional information, are also exploited to improve the constraining power on theoretical models.

In this work, we propose a new probe of gravity using the morphology of cosmic large-scale structure (LSS) as specified by its four Minkowski functionals (MFs). According to Hadwiger's theorem~\cite{Hadwiger57}, for a spatial pattern in 3D, its morphological properties defined as satisfying motion invariance {\red{(i.e., invariant under rotation and translation)}}, {\red{Minkowski}} additivity {\red{(i.e., the property of a union of domains is the sum of those of the individual domains minus that of the overlapping domain)}} etc. are completely specified by four MFs. In studying the MFs of LSS, the patterns are most commonly taken to be the excursion sets of a smoothed density field (such as galaxy number density, matter density)~\cite{SchmalzingBuchert97,Hikage+03,Blake+14}, \ie, regions of space with field value above some specified threshold, which is what we adopt in this work, see~\cite{Mecke+94,Wiegand+14} for other choices of patterns. Geometrically, up to a constant multiplicative factor, the four MFs from the zeroth to the third order represent respectively the pattern's volume, its surface's area, integrated mean curvature, and Euler characteristic (or genus \cite{Gott+86,Matsubara03}).

The MFs are complementary to other statistical observables in probing LSS. While the N-point correlation functions in real space or the corresponding poly-spectra in Fourier space probe the statistics of LSS at specific orders, the MFs comprehensively probe all orders of statistics. Compared to the higher order ($n>2$) statistics whose measurements are usually cumbersome to obtain, the MFs can be easily measured. Moreover, the MFs have the advantage of being more robust to systematic effects such as non-linear gravitational evolution and biases of LSS tracers \cite{Matsubara03,Blake+14}, compared to the N-point statistics. For example, on linear scales, the MFs measured with different tracers of LSS are all the same, \ie, independent on tracer biases, which is an advantageous feature that N-point statistics measurements do not have. 

The MFs were introduced into cosmology by \cite{Mecke+94} in 1994. Since then, most studies on its application have been focused on examining the (non-)Gaussianity of primordial perturbations, from observations of not only the LSS
\cite{[{See e.g. \cite{Hikage+03} and }][{}]Hikage+06,*Codis+13}
but also the CMB \cite{[{See e.g., }][{}]SchmalzingGorski97,*Hikage+08,*Fang+13,*Ade+15,*Buchert+17}, though several other interesting applications have also been proposed \cite{[{See e.g., }][{}]Sato+01,*Park+05,*Gleser+06,*ParkKim10,*Kratochvil+12,*Beisbart+01b}. In particular, its potential in probing gravity was only addressed for the 2D weak lensing convergence field in \cite{Ling+14,*Munshi+16,*Shirasaki+17}. In this paper, motivated by the strong statistical power from ongoing and future LSS surveys, we investigate the potential of these morphological descriptors of LSS in discriminating different theories of gravity. We will base our results on N-body simulations with modified gravity and GR. We notice relevant previous theoretical work by \cite{Wang+12} that used the scaling of the genus statistic of LSS with smoothing scale as a probe of gravity, and relevant discussions by Codis et al in Ref.~\cite{Codis+13}.   

{\it Gravity models and simulations.} As an illustration, we consider two classes of modified gravity models: the Hu-Sawicki $f(R)$ gravity model \cite{hs2007} and the normal-branch Dvali-Gabadadze-Porrati (hereafter nDGP) model \cite{dgp2000,Schmidt09} {\red{(the self-accelerating branch is known to suffer from both observational and theoretical difficulties \cite{Fang+08,*Luty+03,*NicolisRattazzi04,*Koyama07}).}} In each class, we study two models with different parameter values: for $f(R)$ gravity, these are obtained by setting the $f(R)$ parameter $f_{R0}$ to $-10^{-6}$ (F6) and $-10^{-5}$ (F5) respectively, with larger values of $|f_{R0}|$ implying stronger deviations from standard gravity; we tune the nDGP parameter $r_c$ so that the value of $\sigma_8$, the rms linear density perturbation in spherical regions of radius 8$h^{-1}$Mpc at $z=0$, is equal to that in the models of F6 and F5 {\red{(specifically, $r_c=5.7H_0^{-1}$ and $1.2H_0^{-1}$ respectively, with $H_0$ the Hubble parameter)}} -- these are dubbed nDGP$\_$F6 and nDGP$\_$F5 respectively. {\red{Current constraints on $f_{R0}$ and $r_c$ using other cosmological probes are $|f_{R0}|\lsim 10^{-5}$ \cite{Cat+15,*Liu+16,*Pei+17} and $r_c\gsim H_0^{-1}$ \cite{Bar+16} at the $95\%$ C.L., and we expect the MFs of LSS can further tighten these results. Finally, note}} these models are designed to have the same background expansion as the $\Lambda$CDM model.

Our analysis is based on simulations of the above models using the {\sc ecosmog} code \cite{ecosmog,ecosmog-v}. All simulations started at $z=49$ with the same (best-fit WMAP 9yr \cite{wmap9}) cosmological parameters and initial conditions. These are specifically designed to separate effects of the modified gravitational law from those of cosmic variance, background expansion history and other variations in parameters. In all simulations, we evolve 1024$^3$ dark matter particles in a cubic box with length $1024h^{-1}$Mpc a side, which is covered by a regular mesh with 1024$^3$ cells. The cells are refined if they contain more than 8 particles, and such an adaptive refinement scheme \cite{ramses} ensures high force resolution in dense regions, where modified gravity effects are hard to calculate.

{\red{We point out that we have made the hypothesis, following the standard paradigm, that Newtonian simulations for first-order GR perturbations that average out on a FLRW background cosmology faithfully represent structure formation. We call this case ``GR" to distinguish this scenario from corresponding realizations of modified gravity models, the quasi-static approximation used in which is well checked \cite{Bose+15}. 
We emphasize, however, that the standard paradigm is not based on full GR simulations. It may well be that deviations in the form of cosmological backreaction to account for this difference \cite{BuchertRasanen12} could display similar signatures as those of modified gravity models. Our work may imply that backreaction effects could also be quantified by measurement of the MFs.}}

{\it Measurement of Minkowski functionals.} We measure the MFs for the excursion sets of dark matter density field. The procedures are as follows. We obtain dark matter density from positions of particles in our simulations using the cloud-in-cell technique. The density field is subsequently smoothed by a Gaussian window function with size $R_G$. The MFs are then measured for the smoothed field as a function of $\rho$, the density threshold used to define the excursion set. Based on either differential geometry or integral geometry, two standard methods to measure the MFs are developed in \cite{SchmalzingBuchert97}, i.e., the one using the Koenderink invariant and the one using the Crofton's formula. In our calculations, we find the two methods give consistent results. Therefore, we will simply quote our results using the method based on the Crofton's formula.

We denote the four MFs as $V_i$, with $i=0,1,2,3$ specifying their order. Therefore, $V_0$ is the volume fraction of the excursion set, while $V_1,\ V_2,\ V_3$ are its surface's area, integrated mean curvature, and Euler characteristic per unit volume respectively (see e.g. \cite{SchmalzingBuchert97} for exact prefactors in the definitions). The geometrical meaning of $V_3$ may deserve more explanation: it is equal to the genus statistic ($g\equiv$ number of holes minus number of disjoint regions) \cite{Gott+86,Matsubara03} except for a minus sign, therefore, it describes the connectedness of the iso-density contours. $V_3>0$ means more disconnected contours, otherwise, more connected. 

\begin{figure*}[ht]
\hspace*{-0.5cm}
{\includegraphics[angle=0,scale=0.7,trim=0cm 0cm 0cm 0cm,clip=true]{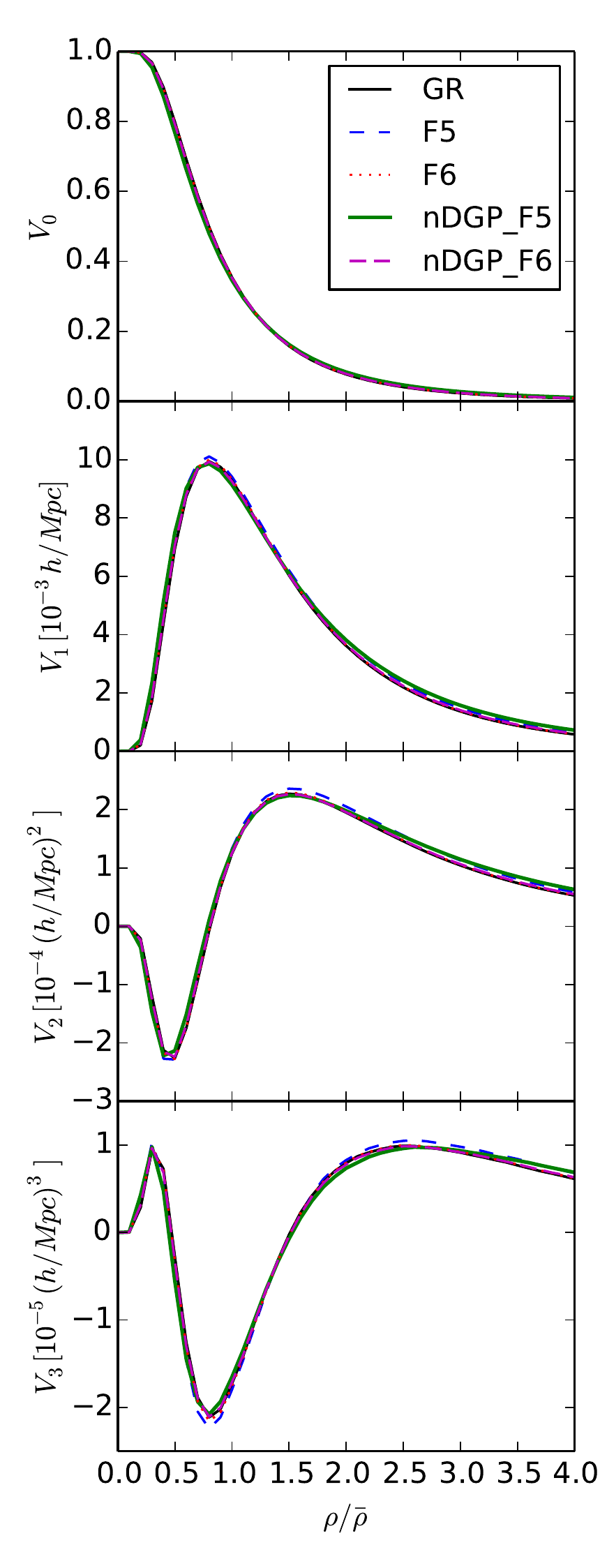}
 \includegraphics[angle=0,scale=0.7,trim=0cm 0cm 0cm 0cm,clip=true]{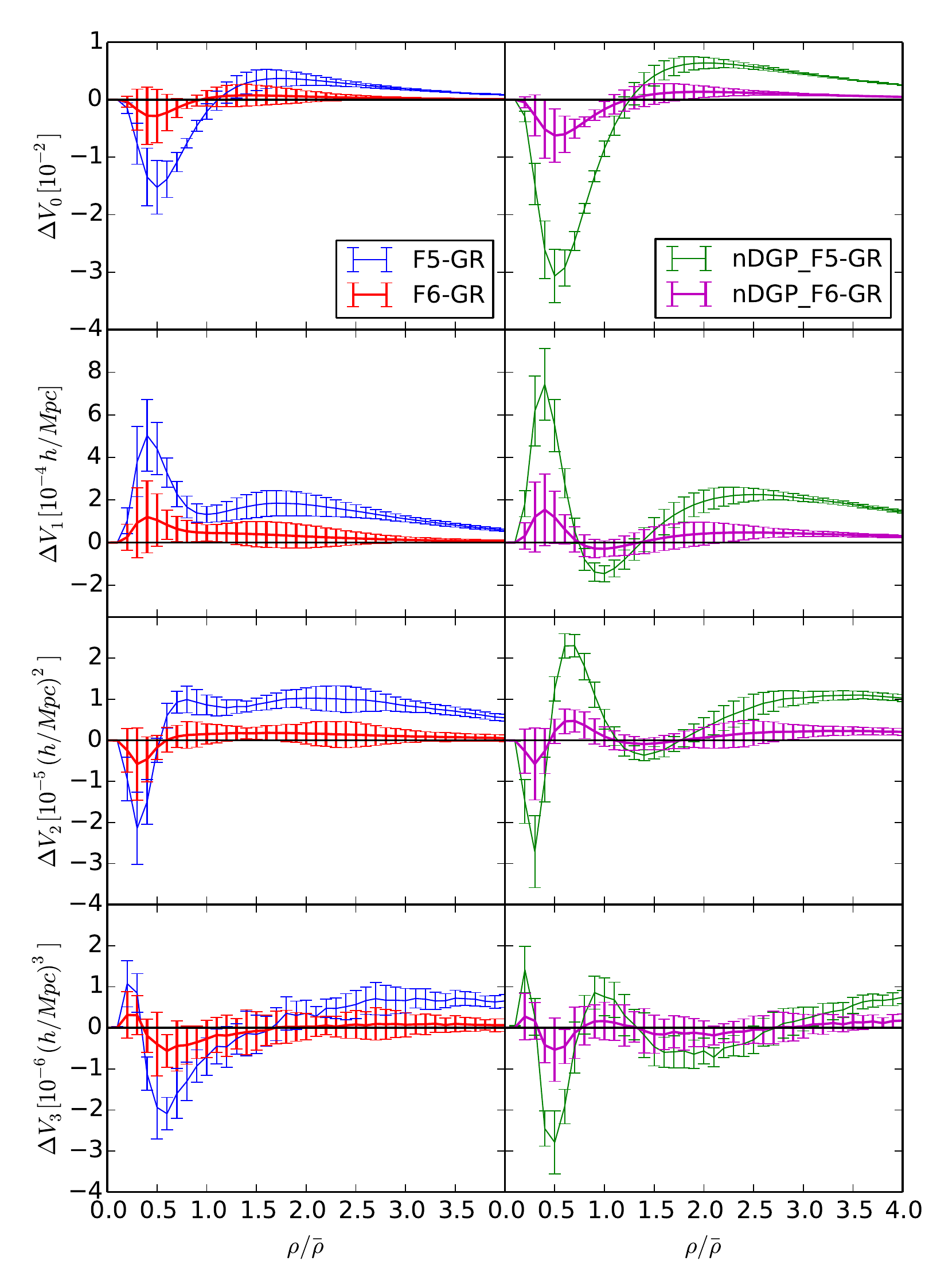}}
\caption{Left panel: The MFs of LSS computed from N-body simulations for different models of gravity at $z=0$ and with $R_G=5h^{-1}$Mpc. The two $f(R)$ models (F5, F6) and two nDGP models (nDGP$\_$F5, nDGP$\_$F6) have the same background expansion as the GR case, i.e. the $\Lambda$CDM model, see text for more details. Right panel: The differences in the MFs between modified gravity models and GR, first column for $f(R)$, and second column for nDGP. $\rho/\bar{\rho}$ is the density threshold used for the MF calculations in ratio of the mean density.
  \label{fig:all} }
\end{figure*}

{\it Results.} To highlight the differences in the MFs caused by modified gravity, we measure and compare the MFs for different models at $z=0$, since the effects of modifications to gravity are generally larger at a lower redshift \cite{ecosmog,ecosmog-v}. We choose the smoothing scale $R_G$ to be 5 $h^{-1}$Mpc, which is large enough to suppress the shot noise without smearing out the important differences in the MFs.

Our results are presented in Figure~\ref{fig:all}. In the left panel, we show the MFs themselves, while the differences in the MFs between modified gravity models and $\Lambda$CDM, the $\Delta V_i$s, are displayed in the right panel. We show the results for $\rho / \bar{\rho}$, ratio of the density threshold to the mean density, in the range of $[0,4]$, over which we find the signal-to-noise ratio (S/N) for the MF is most significant. We estimate the error bars by subdividing our original simulation box into eight equal-sized sub-boxes, each with a volume $512^3h^{-3}$Mpc$^3$ (for small error bars and for convenience of the calculation), and taking the standard deviations of the MFs measured for each sub-box. These are displayed in the right panel of Figure~\ref{fig:all}. 
{\red{We note here that the idealistic way to estimate the error bars is by running a large ensemble of simulations (a highly computationally expensive task which we postpone for future work), and subdividing the simulation box is only indicative for the idealistic error estimation.}}
In the following, we focus on understanding the $\Delta V_i$s. The curves of the MFs themselves share similar trends as a Gaussian random field, which have been well studied in the literature, see e.g. \cite{Blake+14}. Therefore, in the following, they are mentioned only for the purpose of helping to understand the $\Delta V_i$s.  

Overall, we find the differences in the MFs caused by the two $f(R)$ models have the same trend. However, as the modification to gravity is stronger in F5, $\Delta V_i$ in this model has a higher amplitude. The same results hold true for the two nDGP models. While compared with the $f(R)$ model, $\Delta V_i$ in the corresponding nDGP model that has the same $\sigma_8$ as $f(R)$ also has a higher amplitude and more features, though shares a similar trend. This is because the $\sigma_8$ values for $f(R)$ gravity and nDGP are calculated using linear perturbation theory; in $f(R)$ gravity, this overestimates the modified gravity effect by neglecting the chameleon screening, so that the effect in a realistic universe, fully captured by the nonlinear simulations, is actually smaller than suggested by the value of $\sigma_8$; while for nDGP, the screening is weak on linear scales, so that the $\sigma_8$ value is a reasonably accurate description of the realistic case.

Specifically, for $V_0$, {\red{the volume fraction occupied by regions whose densities are above a given threshold,}} we find both the modified gravity in $f(R)$ and nDGP make it larger when {\red{the threshold}} $\rho/\bar{\rho}\gsim 1$ and smaller when $\rho/\bar{\rho}\lsim 1$. That is, the volume fraction with {\red{density}} above an overdensity threshold gets larger, while above an underdensity threshold gets smaller. The latter is equivalent to that the volume fraction with {\red{density}} below an underdensity threshold gets larger. These findings are consistent with the picture that both haloes and voids are more abundant and/or bigger in the two models \cite{lzk2012}. One can also infer that this trend is stronger in nDGP than in the corresponding $f(R)$ model for the same reason as mentioned above, as one finds in Figure~\ref{fig:all}.

For {\red{the excursion sets' surface area}} $V_1$, its difference caused by the two kinds of modification to gravity roughly follows that of {\red{their volume fraction}} $V_0$, except at the very low density threshold region ($\rho/\bar{\rho} \lsim 1$ for $f(R)$ and $\lsim 0.8$ for nDGP) where the changes are in opposite directions. This can be understood as {\red{follows}}: if the excursion sets are all isolated regions, as is the case for high enough density threshold, it is reasonable to expect that the change in their surface area follows that in the volume fraction they occupy. However, for very low density threshold, though the iso-density contours are still isolated, their enclosed regions are no longer the excursion sets, but under-dense regions with {\red{density}} below the threshold, whose volume fraction is therefore $(1-V_0)$. Thus, at the very low density threshold region, $V_1$ changes in the opposite direction as $V_0$, and becomes larger in both $f(R)$ and nDGP. 

For $V_2$, {\red{the integration of the mean curvature of the excursion sets' surface over the surface area,}} we find it to be negative (positive) when $\rho/\bar{\rho}\lsim 0.8(\gsim 0.8)$ in all the models we study. This is understandable considering that positive direction of the surface points toward lower density regions. {\red{Specifically, when the density threshold is high enough, the excursion sets will be isolated high density regions with closed surfaces whose positive directions point outward, therefore the mean curvature is positive; while when the density threshold is low enough, the excursion sets will be the complement of isolated low density regions with closed surfaces whose positive directions now point inward, therefore the mean curvature is negative. The transition from positive to negative of the integrated mean curvature happens at $\rho/\bar{\rho}\simeq0.8$ in the models we study.}} Suppose the change in the mean curvature is negligible, one can expect that $V_2$ changes in the opposite (same) direction as $V_1$ when $\rho/\bar{\rho}\lsim 0.8(\gsim 0.8)$, since the surface area is always positive (note, $V_2$ is roughly the mean curvature times the area). This is roughly the case for $\Delta V_2$ when $\rho/\bar{\rho}$ is far enough from $0.8$ in both the $f(R)$ and nDGP models. While, when $\rho/\bar{\rho}$ approaches $0.8$ and where this phenomenon does not hold, the change in the mean curvature plays a more important role, and we find $V_2$ gets larger in both models. Combined with the specific changes in $V_1$, one can infer that the mean curvature gets larger for the iso-density contours specified with these density thresholds.

As for the connectedness {\red{evaluated by comparing the number of holes through the structure of the excursion sets and the number of disjoint parts in it}}, we find the iso-density contours are more connected with $V_3<0$ {\red{(i.e., more holes)}} when $0.5\lsim\rho/\bar{\rho}\lsim1.5$, but more isolated with $V_3>0$ {\red{(i.e., more disjoint parts)}} elsewhere. For $f(R)$, we find the modification to gravity makes this behavior more evident, \ie, $V_3$ is larger (smaller) where it is positive (negative). This is also roughly the case for nDGP, except that $V_3$ gets larger around $\rho/\bar{\rho}\simeq0.8$ where it was most negative, that is, the contours are less connected where they were most connected. This difference highlights the different modifications to gravity in the two models, and together with other differences in the $\Delta V_i$'s, can be used to discriminate one from the other.

With our estimation for the error bars, we find that the MFs can easily discriminate the modified gravity models of nDGP$\_$F5, nDGP$\_$F6 and F5 from GR: with an individual MF measurement, these models can be discriminated from GR with a significance level up to $\sim$ $30$, $10$ and $5\sigma$. Though for F6, individual MF measurement can only tell it from GR at the level of 1$\sigma$ at most, combining the different orders of MFs at different thresholds can probably boost the level to be significant enough. We also note that the error bars we obtained from simulations are appropriate for a galaxy survey with survey volume of 0.125$h^{-3}$Gpc$^3$ (corresponding to observing the full sky from $z=0$ to $z\simeq 0.1$) and with galaxy number density of $1h^{3}$Mpc$^{-3}$, which are more optimistic compared to the galaxy surveys nowadays. However, considering the strong statistical power we have found and considering the gains from combining the MFs of different orders, at different thresholds, with different smoothing scales and at different redshifts, we expect the MFs can provide a powerful probe of gravity, especially for those ambitious galaxy surveys with large survey volumes and high galaxy densities such as DESI \cite{DESI}. We leave such a comprehensive study with real surveys for future work \cite{Fang+17}.

{\it Conclusions.} The morphological properties of LSS described by the four MFs provide important complementary information to other statistics. In this paper, we investigate their potential as a new probe of gravity. By using N-body simulations of modified gravity and GR, we disclose the morphological differences in LSS caused by modified gravity, and provide insights to understand the differences in the MFs. With the estimated errors for a galaxy survey with volume of 0.125$h^{-3}$Gpc$^3$ and galaxy number density of $1h^{3}$Mpc$^{-3}$, we find the modified gravity models of nDGP$\_$F5, nDGP$\_$F6 and F5 can be discriminated from GR with a significance level up to $\sim$ $30$, $10$ and $5\sigma$ with just an individual MF measurement. This strong statistical power of the MFs as a probe of gravity can probably survive in ambitious real galaxy surveys with smaller survey areas and lower galaxy densities, providing the MFs of different orders, at different thresholds, with different smoothing scales and at different redshifts are combined together. We pursue such a comprehensive study in an ongoing work\cite{Fang+17}.

{\it Acknowledgements.} The authors are grateful to two anonymous referees for their valuable comments. We thank Chris Blake, Thomas Buchert for helpful conversations. WF acknowledges Nuhum Arav for hospitality at Virginia Tech. WF is supported by the National Natural Science Foundation of China Grant No. 11643010, 11653002, 11421303. BL is funded by the UK STFC Consolidated Grant No. ST/L00075X/1 and No.~RF040335. GBZ acknowledges support by the 1000 Young Talents program in China and by the Strategic Priority Research Program ``The Emergence of Cosmological Structures'' of the Chinese Academy of Sciences Grant No.~XDB09000000. 
This work used the DiRAC Data Centric system at Durham University,
operated by the Institute for Computational Cosmology on behalf of the
STFC DiRAC HPC Facility (www.dirac.ac.uk). This equipment was funded by
BIS National E-infrastructure capital grant ST/K00042X/1, STFC capital
grants ST/H008519/1 and ST/K00087X/1, STFC DiRAC Operations grant
ST/K003267/1 and Durham University. DiRAC is part of the National
E-Infrastructure.
To access the data used in this paper please send requests to the authors.

\bibliography{mf}

\end{document}